\documentclass[twocolumn,superscriptaddress,amsmath,aps]{revtex4}
\usepackage{graphicx}
\begin{document}
\title{Diversity of reproduction time scale promotes cooperation
in spatial prisoner's dilemma games}
\author{Zhi-Xi Wu}\email{zhi-xi.wu@physics.umu.se}
\affiliation{Department of Physics, Ume{\aa}\, University, 901 87
Ume{\aa}, Sweden}
\author{Zhihai Rong}\email{rongzhh@gmail.com}
\affiliation{Department of Automation, Donghua University, 200051
Shanghai, China}
\author{Petter Holme}\email{petter.holme@physics.umu.se}
\affiliation{Department of Physics, Ume{\aa}\, University, 901 87
Ume{\aa}, Sweden}
\begin{abstract}
We study an evolutionary spatial prisoner's dilemma game where the
fitness of the players is determined by both the payoffs from the
current interaction and their history. We consider the situation
where the selection timescale is slower than the interaction
timescale. This is done by implementing probabilistic reproduction
on an individual level. We observe that both too fast and too slow
reproduction rates hamper the emergence of cooperation. In other
words, there exists an intermediate selection timescale that
maximizes cooperation. Another factor we find to promote
cooperation is a diversity of reproduction timescales.
\end{abstract}
 \maketitle

\section{introduction}
Many biological, social, and economic systems are built and
sustained by cooperating
individuals~\cite{Ballbook,Nowak2006science}. The issue of how
cooperation emerges and persists among selfish individuals is the
topic of game
theory~\cite{Axelrod,Smith1982book,Hofbauer1998book}. The simplest
and most versatile framework for evolutionary game theory is
perhaps the so called symmetric $2\times2$
games~\cite{Herz1994jtb,Hauert2005ajp,Deobeli2005el}. In such
games, agents can employ two behaviors or strategies ---
cooperative or defective --- in their interaction with others to
optimize their fitness. The goal of this branch of game theory is
to establish the conditions for cooperation (or, equivalently,
defection) to spread. An important $2\times2$ game is the
prisoner's dilemma (PD) that describes the situation when
cooperation gives the highest population-level payoff but, in a
short timescale, defection maximizes the expected payoff of an
individual. In well-mixed populations, cooperators cannot
outperform defectors and are doomed to
extinction~\cite{Hofbauer1998book}. There are, however, many
arguably realistic factors that favor cooperation. For example,
one such factor is space --- if the players are embedded in space,
allowed only to interact with their neighbors, cooperation can
spread in the
population~\cite{Nowak1992nature,Szabo1998pre,Hauert2004nature,
  Santos2005prl,Rong2007pre,Szolnoki2008pa}. The microscopic
explanation for this phenomenon is that  cooperators can form
spatial clusters where the boundaries, although exploited by
defectors, protect the cooperators
within~\cite{Nowak1992nature,Szabo1998pre,Hauert2004nature}. On
complex networks --- scale-free networks in particular ---
cooperators have a tendency to occupy high degree nodes, which
stabilizes
cooperation~\cite{Santos2005prl,Rong2007pre,Szolnoki2008pa}. For
detailed surveys of this field, we refer to
Refs.~\cite{Szaboreview,Nowakbook}.  From these examples we learn
that the spatial organization, the interaction patterns, are
important elements behind the emergence of cooperation. This
raises the question: are there also temporal patterns promoting
cooperation? In a recent work~\cite{Chen2008pre}, Chen and Wang
investigated the effects of what they call ``payoff aspiration''
(roughly, the lower limit of the payoff before a player gets
impatient changes strategy). They found that intermediate
aspiration levels maximize cooperation and argue that too high
levels cause too fast strategy changes, while too low levels make
the agents not adaptive enough. These results suggest a more
fundamental principle --- to achieve cooperation, there might
exist an optimal timescale for updating strategies. In the present
paper, we address this hypothesis by studying evolutionary spatial
PD games and by explicitly considering the time scales of
interaction and selection.

One of our basic assumptions is that the selection timescale is
slower than the interaction timescale. This means that even though
each player has a finite lifetime, it can interact many times
before selection occurs. During their lifetimes, agents do not
change their strategies (or phenotype, in a more biological
parlance).

How can we connect the payoff of the game to fitness? The meaning
of payoff is more problem-dependent than fitness --- payoff is
some result of the interaction that is correlated with
reproduction ability~\cite{Hauert2004nature,Doebeli2004science}
and often equated, mathematically, with
fitness~\cite{Hofbauer1998book,Hauert2004nature,Szaboreview}. Some
works using non-standard relationships between payoff and fitness
are Ref.~\cite{Ohtsuki2006nature} (where a background noise is
added to the payoff to obtain fitness) and
Ref.~\cite{Jiang2009arxiv} (studying non-linear fitness dependence
on payoff). Though there are many possible ways to characterize
the relationship between payoff and fitness, in our model we
assume the two quantities are positively correlated (like in
natural selection), and also including an inheritable fitness
component (so called maternal effect~\cite{rasanen}), which can
appear due to social, environmental or epigenetic mechanisms.

In the rest of the paper, we first introduce our model, defining
the fitness and reproduction time scale. Then we present our
numerical and analytic results in detail and relate these to other
studies.

\section{the model}

\subsection{Payoff and fitness}

In the present study we consider evolutionary games where
individuals, located on a two-dimensional square lattice, play PD
games with their closest neighbors. Following common
practice~\cite{Nowak1992nature,Szabo1998pre}, we consider pure
strategies without self-interaction (where one player also plays
with oneself~\cite{Nowak1992nature}), and use the following
parameterization of the PD: if two players both cooperate ($C$)
then both get a ``reward'' of $1$, two defectors ($D$) both obtain
the ``punishment'' $0$ , a $C$ meeting a $D$ receives ``the
sucker's payoff''  $0$, while $D$ acquires ``the temptation to
defect'', $b\in[1,2]$.  For each site $i$ we define its fitness at
generation $t$ as $f_i(t)=af_i(t-1)+g_i-a\,g_i$, where $t>0$ is
the generation of $i$. $g_i$ is the instantaneous payoff from the
game and the parameter $a\in(0,1)$, the \textit{heritability},
sets the balance between the present and past payoff gains --- the
relative importance of a previous generations, or strength of
maternal effects~\cite{rasanen} decays with a factor $a$ per time
step.

\subsection{Reproduction}

Since our main focus is to evaluate how the diversity of
reproduction timescale affects the evolution of cooperation, we
discard the normal setup that strategy updating of the players is
implemented immediately after a round of game~\cite{Szaboreview}.
Instead, we follow the approach of Ref.~\cite{Roca2006prl}, and
assume that the timescale of selection is slower than that of the
interaction. In particular, we update the population by using
probabilistic dynamics: for each focal site $i$, a reproduction
event occurs with probability $p$, whose magnitude therefore
characterizes the timescale of selection. When $p$ is close to
unity, selection is frequent, while for small values of $p$ the
reproduction event is rare and natural selection is slow. It is
easy to see that the average time of selection on each site is
$1/p$. By this simple way, we are also able to introduce diversity
of reproduction timescales.

When strategies are updated, the focal player $i$ and one randomly
selected neighbor, say $j$, compete for creating an offspring at
$i$. The focal site will, depending on their fitnesses, be
occupied by either the offspring of $i$ or $j$. To be specific,
the neighbor $j$ places a new offspring with the same strategy on
the focal site $i$ with a probability
\begin{equation}\label{eq:w}
W_{j\to i}=\frac{1}{1+\exp{[(f_i-f_j)/\kappa]}},
\end{equation}
where $\kappa$ measures the probabilistic selection intensity (or
noise, or temperature); otherwise $i$ is replaced with an offspring of
the previous individual at $i$. A newborn is assigned the initial
fitness $f_i(0)=1$ (the precise value of $f_i(0)$ does not affect our
conclusions). An alternative interpretation is that the individuals
sometimes can start with new strategies taken from a neighbor with
higher fitness than the individual.

\subsection{Effects of noise}

The noise parameter $\kappa$ is known to have a nontrivial effect
on the evolutionary
trajectories~\cite{Szabo2005pre,Perc2006njp,Ren2007pre}. More
specifically, the highest cooperation levels appear in moderate
amounts of noise. (Here we want to point out such optimal behavior
of noise in promoting cooperation is not universal, but dependent
on the topology of the connectivity structure.) Furthermore, in a
very recent paper~\cite{Szabo2009epl}, Szab\'o \emph{et al.}
reported an interesting result that the Darwinian selection of
``temperature" favors values of $\kappa$ close to $0.1$ which is
related to the optimum providing the highest level of cooperation.
In this paper we will not investigate the effects of $\kappa$
further, but set $\kappa$ to a value, $0.01$, that is low enough
for the effects of the game to dominate the effects of noise, and
high enough for selection to occur.

\subsection{Parameter values and simulation technicalities}
We investigate our model with Monte Carlo (MC) simulations. All
runs are on a $100\times100$ square lattice with periodic boundary
conditions. Initially, either a cooperator or a defector, randomly
chosen with equal probability, occupies each site. During one MC
sweep we go through all players once in a random order. The key
observable is the average fraction $F_c$ of cooperators in the
population in the stationary state, which is measured for the last
$10000$ sweeps of the total simulation time $6\times 10^4$. The
results presented below are averages over twenty realizations from
independent initial configurations.

\begin{figure}
\includegraphics[width=0.8\linewidth]{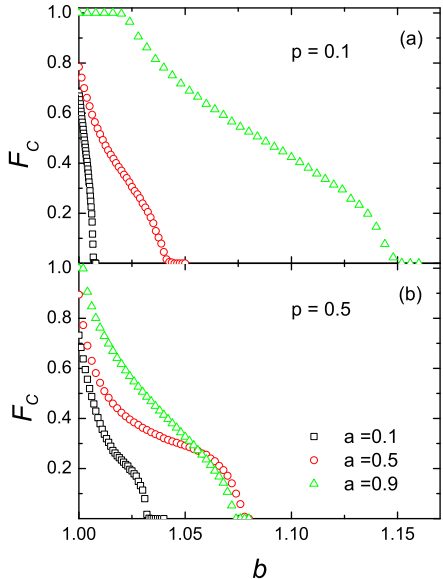}
 \caption{(Color
  online) Average fraction $F_c$ of cooperation as a function of $b$
  for different values of $a$. In (a) and (b) the individuals have
  reproduction rate $p=0.1$ and $0.5$, respectively.} \label{fig1}
\end{figure}

\section{results and analysis}

\subsection{Monomorphic populations}

We start by investigating how the heritability $a$ affects the
evolution of cooperation. The simulation results for $F_c$ as a
function of $b$ for three values of $a$ and two values of $p$ are
shown in Fig.~\ref{fig1}. In general, we can observe that with an
increasing $a$, the average fraction of cooperation grows. For
sufficiently large $a$, the system can even, for the weakest
temptations to defect, reach the absorbing state of full
cooperation. This means that, the ability of inheritance of
fitness from previous generations has a positive impact on the
emergence of cooperation. The larger $a$ is, the greater is the
cooperation level. This result is somewhat expected since other
studies have shown that the average payoff of $C$ in the
equilibrium is greater than that of $D$~\cite{Chen2009}.

\begin{figure}
\includegraphics[width=0.8\linewidth]{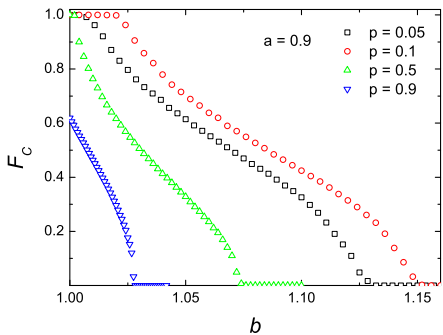} \caption{(Color
  online) Average fraction $F_c$ of cooperators as a function of $b$
  for different values of $p$. The parameter $a$ is fixed to $0.9$.}
\label{fig2}
\end{figure}

\begin{figure}
\includegraphics[width=0.8\linewidth]{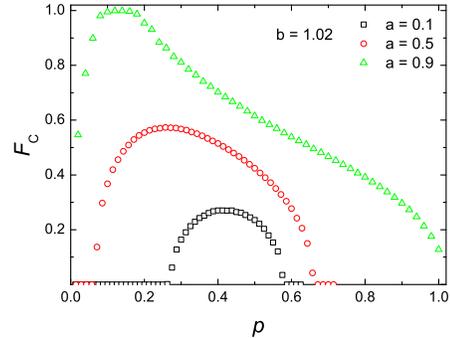} \caption{(Color
  online) Average fraction of cooperation $F_c$ as a function of
  reproduction rate $p$ for $b=1.02$. The squares, circles, and
  triangles correspond to $a=0.1$, $0.5$, and $0.9$,
  respectively.} \label{fig3}
\end{figure}

Even if the $a$-dependence of $F_c$ is qualitatively expected, it
is rather different for the two $p$-values of Fig.~\ref{fig1}(a)
and (b). In Fig.~\ref{fig2} we fix $a=0.9$ and study how the
average fraction of cooperation evolves as $p$ changes. For large
($p=0.9$) and small ($p=0.05$) values $F_c$ is smaller than for
$p=0.1$. This result suggests that a moderate reproduction rate
promotes the spread of cooperation. To make this point clearer, in
Fig.~\ref{fig3} we summarize the results of $F_c$ as a function of
$p$ for different values of $a$. As surmised, the highest
cooperation level is achieved in a region where $p$ is moderate,
resembling coherence resonance phenomenon --- that $p$ alone
(without e.g.\ any weak periodic force) can promote cooperation,
as has been found frequently in recent
literatures~\cite{Perc2006njp,Ren2007pre}. For example, by
introducing additive noise to the payoff matrix of the PD game,
Perc found that cooperation is maximized at a moderate variance of
payoff variations~\cite{Perc2006njp}. Along a different research
line, Ren \textit{et al.}\ studied PD game on random regular
graphs and found that there exists an optimal amount of
topological randomness, leading to the highest level of
cooperation~\cite{Ren2007pre}. Going back to our study, note that
the optimal value $p_o$, where $F_c$ is maximized, depends on the
value $a$. Increasing $a$ will shift $p_o$ toward smaller values
(smaller reproduction rates). For instance, for $a=0.1$ we have
$p_o\approx0.41$, while for $a=0.9$ $p_o$ is about $0.12$. This
point is also reflected by Figs.~\ref{fig1} and \ref{fig2}, where
given $a=0.9$ the curves for $p=0.1$ take the rightmost position.

The nontrivial behavior of $F_c$ in Figs.~\ref{fig1}, \ref{fig2},
and~\ref{fig3} can be qualitatively understood as follows. As $p\to 1$
--- selection happens frequently; the average fitness of $C$ should be
lower than that of $D$ (due to the frequent exploitation on $C$ by
$D$), so $F_c$ will decrease. On the other hand, in case of $p\to
0$ --- selection is very slow; defectors around the boundary have
enough time to obtain a fitness high enough to outcompete the
cooperators. This is reminiscent of the observation that the
formation of $C$-clusters can protect cooperation in the spatial
PD, whereas the clusters of  $D$ would deteriorate their own
environment~\cite{Szabo1998pre,Szabo2005pre,Chen2009}. In the
light of this, the maximal cooperation levels are reached for
$p$-values that are  neither be too large nor too small, so that
$C$ has a larger chance than $D$ to obtain high fitness. In order
to investigate this hypothesis further, we will first check what
will happen if the population is well mixed rather than
two-dimensionally extended. Though mean-field treatment is not
appropriate for spatial case, it is a fair first-order
approximation of the well-mixed case.

For simplicity we assume that the average coordination number
(number of interactions per site, per MC sweep) is $z$. In the
mean-field approximation we can write the following equation for
the evolution of the average fraction of cooperation:
\begin{widetext}
\begin{eqnarray}
\frac{\partial F_c}{\partial t} &=&(1-F_c)F_cW_{D\to
  C}-F_c(1-F_c)W_{C\to D} =F_c(1-F_c)\Big( \frac{1}{1+\exp
  [(\overline{f}_D-\overline{f}_C)/\kappa ]}- \frac{1}{1+\exp
  [(\overline{f}_C-\overline{f}_D)/\kappa ]}\Big)\nonumber\\
&=&-F_c(1-F_c)\tanh{\Big(\frac{\overline{f}_D-\overline{f}_C}
  {2\kappa}\Big)}\label{Fc},
\end{eqnarray}
\end{widetext}
where $\overline{f}_{C,D}$ denote the average fitness of $C$ and
$D$. The sign of $\overline{f}_D -\overline{f}_C$ determines
whether $F_c$ converges to $1$, $0$, or remains unchanged. Since
each individual reproduces with probability $p$, the average
reproduction period is  $T=\lceil1/p\rceil$ where
$\lceil\cdot\rceil$ represent the smallest integer larger than the
argument.  The average fitness of $D$ at age $t$ can be written as
$\overline{f}_D(t)=a\overline{f}_D(t-1)+zbF_c$. Assume the age of
the individuals are equally distributed in the region $t\in[1,T]$,
using the boundary condition $\overline{f}_D(0)=1$, we obtain the
average expected fitness of $D$: $\overline{f}_D=\sum_{i=0}^{T-1}
\overline{f}_D(t)$. After some algebra, we get
\begin{subequations}
\begin{eqnarray}\overline{f}_D &=&
  p\Big(\sum_{i=1}^{T-1}a^i+zbF_c\sum_{i=0}^{T-1}\sum_{j=0}^ia^j\Big),\\
\overline{f}_C &=&
p\Big(\sum_{i=1}^{T-1}a^i+zF_c\sum_{i=0}^{T-1}\sum_{j=0}^ia^j\Big).
\end{eqnarray}
\end{subequations}
We can see that for any $b>1$, because
$\overline{f}_D-\overline{f}_C>0$, $F_c$ will go to zero. The
incorporation of spatial effects or diversity of reproduction
times will change this picture. The permission of only
nearest-neighbor interaction, and non-synchronous reproduction,
also assures the possibility of $\overline{f}_D-\overline{f}_C<0$
at a local spatial scale. Another factor is that $D$ clusters
``defeat themselves''~\cite{Szabo1998pre,Szabo2005pre,Chen2009}.
In combination with these observations, the intrinsic
stochasticity of probabilistic reproduction gives $C$ a chance to
outcompete $D$ for dispersion, given that $p$ is appropriately
selected. As a consequence, cooperators are favored by diversity
of reproduction, and can even dominate the whole population
provided that $b$ is sufficiently small.

\subsection{Polymorphic populations}

In reality there is a spread in reproduction times. We proceed to
investigate the effects of diversity on our PD model. For
simplicity, we restrict ourselves to a bimodal distribution of
$p$-values:  we assign $p=0.1$ to a fraction $v$ of the players
and $p=0.9$ to the rest.

\begin{figure}
\includegraphics[width=0.8\linewidth]{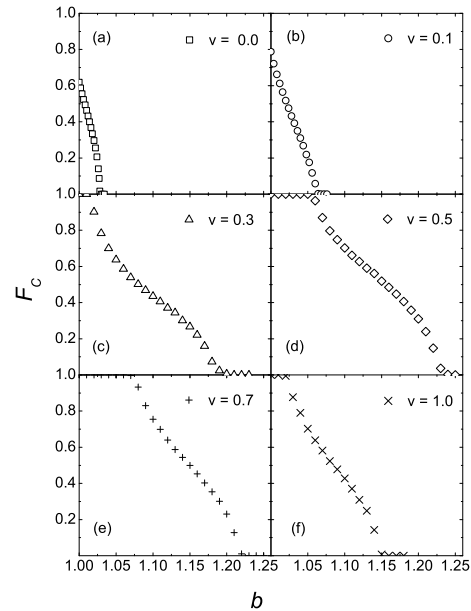}
\caption{Average fraction of cooperation $F_c$ as a function of $b$
  for different systems where fraction $v$ of population has
  reproduction rate $0.1$, and the others have $p=0.9$. The parameter
  $a$ is set to $0.9$.}
\label{fig4}
\end{figure}

\begin{figure}
\includegraphics[width=0.8\linewidth]{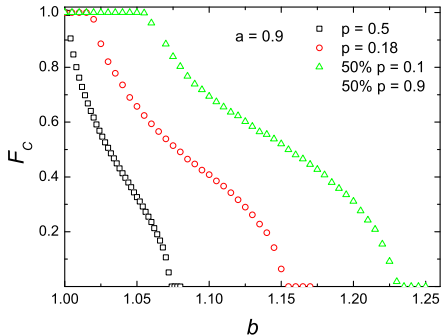}
\caption{(Color online) Average fraction $F_c$ of cooperators as a
  function of $b$ for three special cases where all players have a
  reproduction rate $p=0.5$ (squares), $p=0.18$ (circles), and $50\%$
  individuals with $p=0.1$ and the remaining with $p=0.9$
  (triangles). Note that the population with half $p=0.1$ and half
  $p=0.9$ have the same average reproduction rate as the population with
  $p=0.5$, and the same average life span as the population with $p=0.18$.}
  \label{fig5}
\end{figure}

In Fig.~\ref{fig4} we show the dependence of $F_c$ on $b$ for
$a=0.9$ and different values of $v$. We find that for appropriate
mixing of the two types of individuals, for instance $v=0.5$ and
$0.7$, the final cooperative level is further enhanced as compared
to a monomorphic population ($v=0$ or $1$). To investigate this
further, in Fig.~\ref{fig5} we graph $F_c(b)$ for three systems:
one with a polymorphic population where half of the population has
$p=0.1$ the other has $p=0.9$, another with a monomorphic
population of the same average reproduction rate $p=0.5$ as the
polymorphic population, and a third monomorphic population with
the same average life time as the polymorphic population (giving
$p=0.18$). In this comparison, the polymorphic population has a
higher cooperation level than both the monomorphic reference
populations.

\begin{figure}
\includegraphics[width=0.8\linewidth]{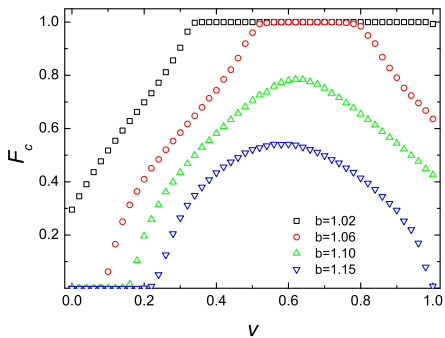}
\caption{(Color online) Average fraction of cooperators as a function
  of the balance between reproduction rates $v$ for different values
  of $b$. $v=0$ or $1$ represent monomorphic populations; diversity
  (measured in standard deviation) is largest for $v=1/2$. $a$ is set
  to parameter $0.9$.}
\label{fig6}
\end{figure}

From Fig.~\ref{fig4}, we also note that an intermediate mixing of
reproduction rate of the population benefits the maintenance of
cooperation.  This finding is corroborated by Fig.~\ref{fig6}
where $F_c(v)$ is displayed for various $b$-values. For small $b$,
$F_c$ can reach unity even for moderate values of $v$. With an
increasing temptation to defect, the region of $v$ with non-zero
cooperation vanishes. Despite this, the peak of $F_c$ is located
at intermediate $p$-values.

\begin{figure}
\includegraphics[width=0.8\linewidth]{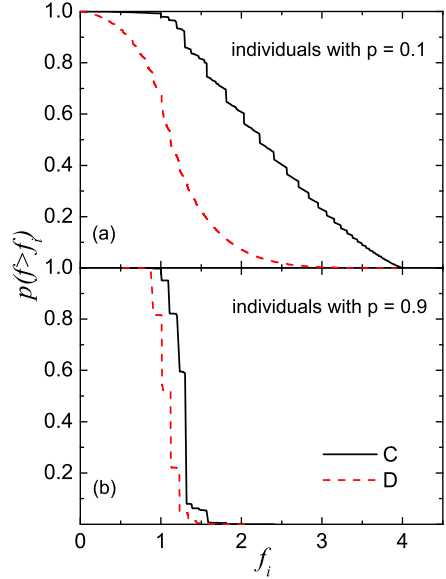}
\caption{(Color online) The cumulative distribution of the fitness of
  agents with different $p$-values in a mix of half $p=0.1$, half
  $p=0.9$. (a) shows the fitness for agents with $p=0.1$, (b) shows
  curves for agents with $p=0.9$. Solid and dashed lines are for $C$,
  $D$, respectively. Other parameters: $a=0.9$, $b=1.1$. Under such
  parameterization, the final fraction of $C$ with $p=0.1$, and
  $p=0.9$, is about 0.38(2), and 0.33(2); the corresponding $D$ with
  $p=0.1$, and $p=0.9$, is about 0.12(2), and 0.17(2), respectively.}
\label{fig7}
\end{figure}

The enhancement of cooperation at intermediate levels of mixing of
two types of individuals can be explained in the same vein as
Ref.~\cite{Szolnoki2007epl}, where Szolnoki and Szab\'o studied
evolutionary PD game with a fraction $v$ of players having higher
influence to spread their strategies than the others. They found
that highest level of cooperation is attained when $v$ is around
$0.5$. Their interpretation was that the less influential
individuals act as followers~\cite{Szolnoki2007epl} which
stabilizes cooperation. In our case, by analogy, we can assume
individuals with high reproduction (large $p$) act as
followers~\cite{Szolnoki2007epl}, since, on average, they always
get lower fitness than those with slow reproduction rate. This
point can be verified by the data presented in Fig.~\ref{fig7},
where the cumulative distribution of the fitness of the population
is shown. In this sense, even though the microscopic mechanisms
are different, both our present model and that of
Ref.~\cite{Szolnoki2007epl} can be regarded as belonging to a
generalized class that adjusts the temporal organization of the
individuals in a way that maximizes cooperation.

\begin{figure}
\includegraphics[width=0.8\linewidth]{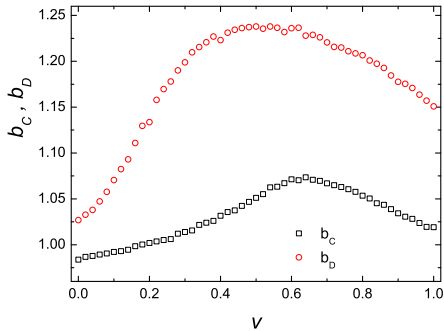}
\caption{(Color online) The extinction thresholds $b_C$, $b_D$
  of cooperators and defectors, respectively, as a function of
  $v$. The parameter $a=0.9$. Note that the range of $b$ extends to
  values lower than 1, where the game is not a proper prisoner's
  dilemma.}
\label{fig8}
\end{figure}

To further investigate how the difference of reproduction rate
affects the evolution of cooperation; we measure the
$v$-dependence of extinction threshold of $b_D$ and $b_C$ --- the
values of $b$ that demarcate the region of coexisting cooperators
and defectors. The simulation results are shown in
Fig.~\ref{fig8}. Between the upper and lower critical points $C$
and $D$ can coexist. Above (below) this region only defectors
(cooperators) remain alive in the final stationary states. Note
that, for the sake of comparison, we extend the $b$-values to the
region smaller than $1$ where the game is a Stag-Hunt game rather
than a PD~\cite{Szaboreview}. From this figure, we observe that
for intermediate $v$ both $b_D$ and $b_C$ attain their maximum
values, though not at the same $v$.

\section{conclusions and discussion}
To summarize, in the present work we have proposed and analyzed an
evolutionary, spatial prisoner's dilemma game including a partial
inheritance of fitness, a probabilistic reproduction, and a
diversity of reproduction rates. The combination of these factors
gives rise to rich dynamic behavior of the system. In particular,
the stronger the ability of inheritance of fitness is, the more
cooperation is there. The reproduction rate has a nontrivial
effect on the evolution of cooperation. When the parameter $a$
(characterizing the strength of inheritance of fitness) is fixed,
the highest cooperation level is realized for moderate
reproduction rates. Both too slow and too fast reproduction rates
can harm cooperation. Moreover, we find that the presence of
diversity of reproduction rate can enhance cooperation. The
behavior of promoting cooperation is found to resemble coherence
resonance. These findings suggest that the heterogeneity of
individual traits might benefit cooperation in PD situation,
giving another clue to the emergence of population in social and
biological systems of selfish individuals.

In a recent work, Qin \textit{et al.}~\cite{Qin2008pre} explored
the effects of infinite memory in a spatial PD game (allowing
self-interaction). They found that with a moderate memory effect,
the extinction threshold $b_D$ reaches a maximal value, but
observed no such behavior for $b_C$. Their memory mechanism is
similar to the maternal-effect fitness inheritance in our model.
In another related work~\cite{Roca2006prl}, Roca \textit{et al.}\
considered independent interaction and selection timescales in
evolutionary games. The authors studied all $12$ nonequivalent
symmetric $2\times2$ games, and found that, depending on the type
of the game, rapid selection may lead to changes of the asymptotic
equilibria and changes in the basins of
attraction~\cite{Roca2006prl}. Ref.~\cite{Roca2006prl} differs
from ours in the approach to strategy updates --- they model
updates as a Moran process, whereas we use a stochastic migration
rule; and in the interaction configuration they, in contrast to
us, consider a non-spatial (or well-mixed) population.

One future problem is how cooperators vanish as the temptation to
defect increases. For evolutionary PD with stochastic imitation
rules Szab\'o and T\H{o}ke~\cite{Szabo1998pre} found a phase
transition of the  directed percolation universality
class~\cite{Szabo1998pre}. We expect a similar transition can
occur in our model in the case of large value of $p$. However,
this point needs to be confirmed in further studies. Moreover, how
the $p$-value maximizing $F_c$ depends on the parameter $a$ is not
addressed in the present work. The work in~\cite{Santos2005prl}
suggests that large degree nodes are important for promoting
cooperation. In a recent paper~\cite{Chen2009pre}, middle degree
nodes, on the other hand, are found to be crucial for stabilizing
cooperation. Along these lines, we expect that the integration of
inhomogeneous interaction pattern and reproductive diversity to
our present model would be interesting.

\begin{acknowledgements}
The authors thank Etsuko Nonaka for comments. Z.X.W. acknowledges
support from the Swedish Research Council. H.R. acknowledges
support from the National Science Foundation of P. R. China (No.\
60874089). P.H. acknowledges support from the Swedish Foundation
for Strategic Research.
\end{acknowledgements}

\bibliographystyle{h-physrev3}

\end{document}